\documentclass[11pt,a4paper]{article}
\pdfoutput=1

\usepackage{jheppubnohead}
\usepackage[utf8]{inputenc}
\usepackage{amsmath}
\usepackage{epsfig}
\usepackage{graphicx}
\usepackage{amssymb}
\usepackage{tabularx}
\usepackage[normalem]{ulem} 
\usepackage{booktabs} 
 
\def\beqn{\begin{eqnarray}}
\def\eeqn{\end{eqnarray}} 
\def\be{\begin{equation}}
\def\ee{\end{equation}}
\def\nn{\nonumber}

\def\mc{\mathcal}

\renewcommand{\d}[1]{\ensuremath{\operatorname{d}\!{#1}}}

 \title{Closing in on singlet scalar dark matter: LUX, invisible Higgs decays and gamma-ray lines}
 
 \author[a]{Lei~Feng,}
 \author[b]{Stefano~Profumo,}
 \author[c]{Lorenzo~Ubaldi}

 \affiliation[a]{Key Laboratory of Dark Matter and Space Astronomy, \\
Purple Mountain Observatory, Chinese Academy of Sciences, Nanjing 210008, China}
\affiliation[b]{Department of Physics and Santa Cruz Institute for Particle Physics \\
University of California, Santa Cruz, CA 95064, USA}
\affiliation[c]{Raymond and Beverly Sackler School of Physics and Astronomy, \\
 Tel-Aviv University, Tel-Aviv 69978, Israel}

\emailAdd{fenglei@pmo.ac.cn}
\emailAdd{profumo@ucsc.edu}
\emailAdd{ubaldi.physics@gmail.com}

\abstract{We study the implications of the Higgs discovery and of recent results from dark matter searches on real singlet scalar dark matter. The phenomenology of the model is defined by only two parameters, the singlet scalar mass $m_S$ and the quartic coupling $a_2$ between the SU(2) Higgs and the singlet scalar. 
We concentrate on the window $5 < m_S/{\rm GeV} < 300$. The most dramatic impact on the viable parameter space of the model comes from direct dark matter searches with LUX, and, for very low masses in the few GeV range, from constraints from the invisible decay width of the Higgs. In the resonant region the best constraints come from gamma-ray line searches. We show that they leave only a small region of viable parameter space, for dark matter masses within a few percent of half the mass of the Higgs. We demonstrate that direct and indirect dark matter searches (especially the search for monochromatic gamma-ray lines) will play a key role in closing the residual parameter space in the near future.
}

\begin{document}

\maketitle

\section{Introduction}
One of the arguably most economical extensions to the Standard Model (SM) of particle physics that provide a dark matter (DM) candidate consists of augmenting the SM with a real gauge-singlet scalar $S$ charged under a global $Z_2$ symmetry under which $S$ is charged ($S \to -S$)
and all other SM fields are neutral. $S$ is then a prototypical weakly interacting massive particle (WIMP) that interacts with other SM fields via mixing with the SU(2) Higgs.

Such a minimal extension to the SM has a long history. A first incarnation of a singlet, real scalar extension to the SM was envisioned by Veltman and Yndurain~\cite{Veltman:1989vw} in the context of one-loop radiative corrections to SM processes such as $WW$ scattering. The scalar particle was first considered in a ``cosmological'' context, to our knowledge, by Silveira and Zee~\cite{Silveira:1985rk}, where the relic abundance from thermal freeze-out for a stable real scalar gauge singlet was first calculated; Ref.~\cite{Silveira:1985rk} also computed the ``direct detection'' cross section, i.e. the scattering cross section for the scalar particle off of baryons, and other quantities relevant for the phenomenology of the mode, such as the impact on the SM Higgs decay and on the flux of Galactic cosmic rays.

Later incremental work on this setup included Ref.~\cite{McDonald:1993ex}, which considered an arbitrary number of complex singlet scalars, and Ref.~\cite{Burgess:2000yq}, which focused on collider implications, on possible DM self-interactions effects, and on constraints from the singlet potential of the model. Subsequent studies that focused on the phenomenology of a singlet scalar extension to the SM at colliders and with DM searches include Refs.~\cite{Barger:2007im, Damgaard:2013kva, No:2013wsa, He:2009yd, Gonderinger:2009jp, Cline:2013gha, Petraki:2007gq, Queiroz:2014yna, Drozd:2011aa, Djouadi:2011aa, Djouadi:2012zc, Raidal:2011xk, Baek:2014kna, Baek:2014jga, Ko:2014gha, Stojkovic:2013ppa, Robens:2015gla, Cheung:2012xb, Yaguna:2011qn, Goudelis:2009zz, deSimone:2014pda}. Not long ago, two of us (SP and LU) have focused~\cite{Profumo:2010kp} on the issue of vacuum stability in this model~\cite{Kadastik:2011aa}, as well as on a first calculation of the pair-annihilation cross section of the additional singlet into two photons.

The possibility that this simple extension to the SM might be relevant for models where the baryon asymmetry is produced at the electro-weak phase transition has also been widely explored~\cite{Espinosa:1993bs, Noble:2007kk,   Cline:2009sn,  Cline:2012hg, Katz:2014bha, Fuyuto:2014yia}. A strongly first-order electroweak phase transition is in fact a generic possibility that this model entails (see e.g. the recent analysis of Ref.~\cite{Profumo:2014opa} and references therein).

There are two reasons why it is now timely to scrutinize the singlet real scalar DM model:
\begin{itemize}
\item[{(i)}] The discovery of a 125 GeV Higgs by the CMS~\cite{Chatrchyan:2012ufa} and ATLAS~\cite{Aad:2012tfa} Collaborations at the Large Hadron Collider, a discovery which effectively removes one degree of freedom from the model parameter space, and
\item[{(ii)}] Recent, rapid progress in the area of both direct and indirect dark matter detection, with significant improvements on constraints on the size of the spin-independent scattering cross section of DM particles off of nuclei~\cite{Akerib:2013tjd} and on the pair-annihilation cross section of DM into two monochromatic gamma-ray photons~\cite{Ackermann:2013uma}.
\end{itemize}
As we show in the present study, the singlet real scalar DM model {\em is alive}, but the only viable region of parameter space of this model where the DM can be produced as a thermal relic from the early universe is  highly constrained and will be thoroughly explored in the very near future.

The three key novelties we bring with the present study are: (i) a comprehensive and updated overview of all relevant available direct, indirect and collider searches constraints on the model under consideration; (ii) a new calculation of the $\gamma Z$ line in the model and an update on the Fermi gamma-ray line constraints which we show are key to probe the resonant region, effectively the only region in the model with relatively low particle masses which was not yet excluded; and (iii) we provide an accurate calculation of the thermal relic density of the dark matter candidate in this model, with minor but significant differences from previous, less accurate calculations.

We introduce the model and the notation in the following Section \ref{sec:model}, and present our results in Sec.~\ref{sec:results}. We also include two appendices: in the first one we list the relevant expressions for the pair-annihilation cross sections used to derive our constraints, while in the the second one we give details on the derivations of the Fermi constraints on the singlet annihilation into $\gamma Z$.  
 
 \section{Model Setup}\label{sec:model}
 
The Lagrangian for the model we consider here is defined by the following expression:
\be
{\mc L} ={\mc L}_{\rm SM}+\frac{1}{2}\partial_\mu S \partial^\mu S-\frac{b_2}{2} S^2-\frac{b_4}{4} S^4-a_2 S^2
H^\dagger H \, ,
\ee
where ${\mc L}_{\rm SM}$ is the SM Lagrangian, $H$ is the SM Higgs doublet, and we are using the notation of 
Refs.~\cite{Profumo:2007wc, Profumo:2010kp}. We require that the Higgs gets a non-vanishing vacuum expectation value (VEV) $v=246$ GeV at the
minimum of the potential while the
singlet does not, $\langle S \rangle = 0$, to ensure stability of the DM candidate \cite{Profumo:2007wc}. After electroweak symmetry breaking, writing $H^\dagger = 1/\sqrt{2} (h+v,0)$ with $h$ real, the scalar potential reads
\be \label{eq:potential}
V(h,S)=-\frac{\mu^4}{4\lambda}-\mu^2 h^2+\lambda v h^3+\frac{\lambda}{4} h^4+\frac{1}{2}(b_2+a_2 v^2)S^2+\frac{b_4}{4} S^4+a_2 v S^2 h+\frac{a_2}{2}S^2 h^2 \, ,
\ee
where $\mu^2<0$, $\lambda$ is the quartic coupling for the Higgs, and $(-\mu^2/\lambda)^{1/2}=v$. This potential is bounded from below, at tree level, provided that $\lambda, b_4 \geq 0$, and $\lambda b_4\geq a^2_2$ for negative $a_2$. The singlet mass is, at tree level,
\be \label{eq:mS}
m^2_S = b_2 + a_2 v^2 \, .	
\ee
The phenomenology of this model is completely determined by the parameters $a_2$ and $b_2$ (or $m_S$),
since the self-interaction quartic coupling $b_4$ does not play any phenomenologically observable role (see e.g. \cite{Profumo:2007wc, Profumo:2010kp}).

In this paper we study experimental bounds on the two-dimensional parameter space $\{ a_2, m_S \}$ and we 
update the results of our previous work~\cite{Profumo:2010kp}. Since then, the Higgs has been discovered~\cite{Aad:2012tfa, Chatrchyan:2012ufa}, 
thus its mass is no longer a free parameter. In addition, we also now have constraints on the invisible Higgs decay $h \to SS$~\cite{Zhou:2014dba, Aad:2014iia, Chatrchyan:2014tja}, and both direct~\cite{Akerib:2013tjd, Akerib:2015rjg} and indirect~\cite{Ackermann:2013uma} detection limits have improved  significantly. 
  
  \begin{figure}[t]
\centering
\includegraphics[width=0.8\textwidth]{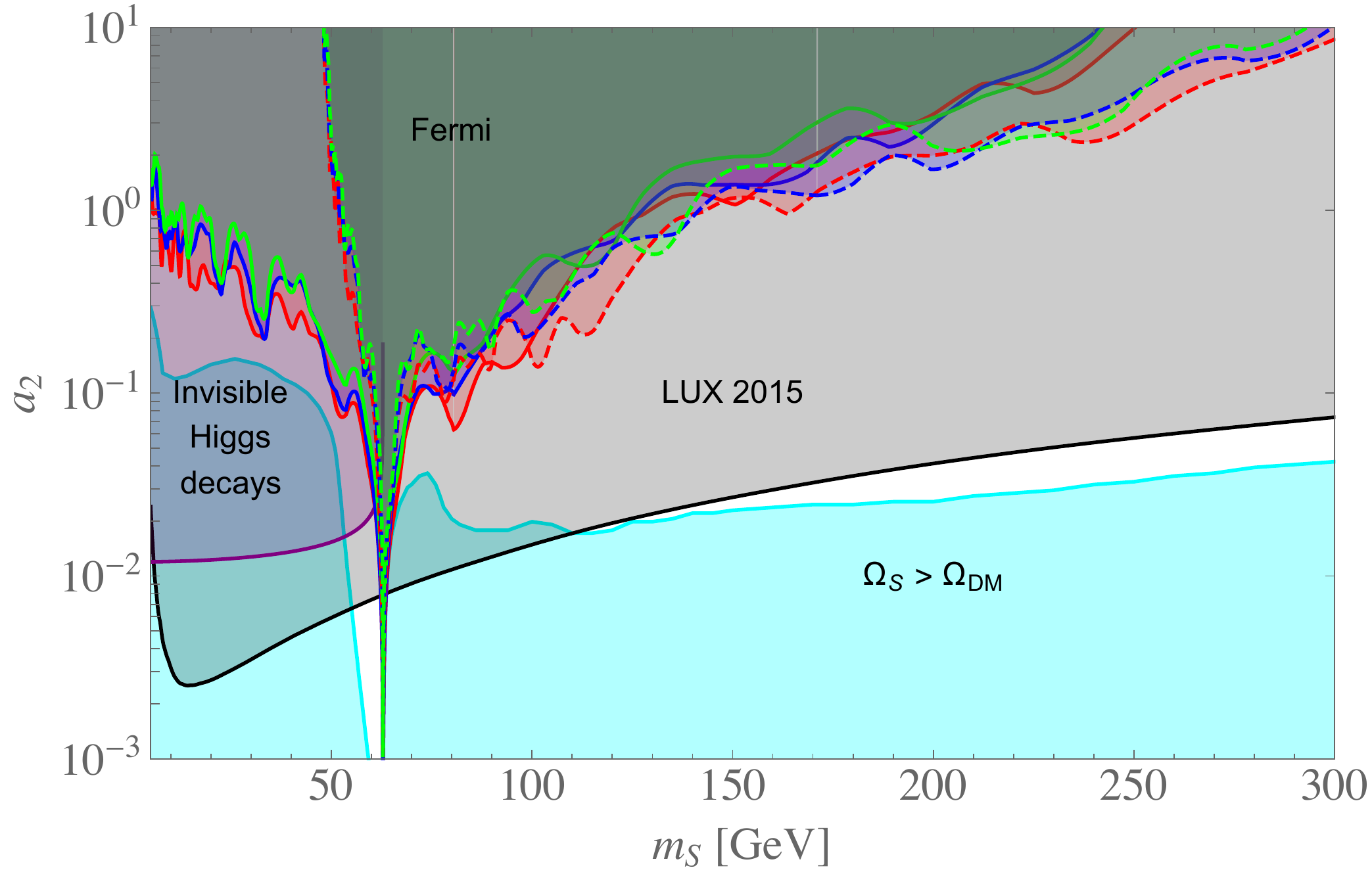}
\caption{Along the cyan line the real scalar singlet gives the correct dark matter relic abundance. The region below this line corresponds to overabundance and is excluded, while most of the
region above is excluded by experimental constraints. The strongest limits are from direct detection (LUX~\cite{Akerib:2015rjg}): they exclude
the region above the black line. Going to masses below a few GeV the most important constraint comes from invisible Higgs decays searches~\cite{Zhou:2014dba}, which exclude the region above the purple line. We show several lines for the constraints from gamma-ray line searches (Fermi~\cite{Ackermann:2013uma}): the plain 
lines correspond to the annihilation $SS \to \gamma \gamma$, the dashed lines to $SS \to \gamma Z$. The colors
correspond to different dark matter density profiles: red is for Einasto, blue for NFW, green for Isothermal. Fermi excludes the area above these lines. The only regions which are not yet excluded are the white areas, one for $m_S > 110$ GeV, the other on the lower
left part of the plot, close to the resonance $m_S = m_h/2$.
We zoom into the resonant region in Fig.~\ref{fig:resonantregion}. }
\label{fig:all}
\end{figure}

\section{Results}\label{sec:results}
  Our results are summarized in Fig.~\ref{fig:all}. The cyan line in the plot represents the region of parameter space where we obtain the correct dark matter relic abundance for $S$. To compute the relic density we solve numerically
  the Boltzmann equation\footnote{We follow here the notation of Ref.~\cite{Cline:2013gha}.}
 \be \label{eq:BE}
 \frac{\d Y}{\d x} = Z(x) \left[Y^2_{\rm eq}(x) - Y^2(x) \right] \, ,
  \ee
  where $Y \equiv n/s$, with $n$ the number density of the scalar $S$, $s$ the entropy density, $x \equiv m_S/T$,
  \beqn
  Z(x) &=& \sqrt{\frac{\pi}{45}} \frac{m_S M_{\rm Pl}}{x^2} [\sqrt{g_*} \langle \sigma v_{\rm rel} \rangle ] (x) \, , \label{eq:Z}\\
  Y_{\rm eq}(x) &=& \frac{45}{4 \pi^4} \frac{x^2}{h_{\rm eff} (x)} K_2 (x) \, , \\
  \sqrt{g_*} &=& \frac{h_{\rm eff}}{\sqrt{g_{\rm eff}}} \left( 1 + \frac{T}{3 h_{\rm eff}} \frac{\d h_{\rm eff}}{\d T} \right) \, .
  \eeqn
  Here $T$ is the temperature, $M_{\rm Pl}$ the reduced Planck mass, $h_{\rm eff}$ and $g_{\rm eff}$ the effective
  entropy and energy degrees of freedom, computed assuming SM particle content, $K_2(x)$ a modified Bessel function of the second kind. Eq.~\eqref{eq:BE} is the usual Boltzmann equation that one has to solve in order to find the relic density of a WIMP. 
  Particular attention has to be paid to the thermal averaged annihilation cross section, especially in the resonant region.
  We follow the prescription of Ref.~\cite{Gondolo:1990dk}
  \be \label{eq:GG}
  \langle \sigma v_{\rm rel} \rangle = \int_{4 m_S^2}^\infty \frac{s \sqrt{s-4m_S^2} K_1(\sqrt{s}/T) \sigma v_{\rm rel}} {16 \ T \ m_S^4 \ K_2^2(m_S/T)} \d s \, .
  \ee
  Here $K_1 (\sqrt{s} / T)$ is a modified Bessel function of the second kind, and $s$ is the square of the center-of-mass energy. The details of the calculation of $\sigma v_{\rm rel}$, which appears in the integral of eq.~\eqref{eq:GG}, are in Appendix~\ref{sec:cross}.

 Most of the region in which $S$ would give the observed dark matter abundance is
  ruled out by LUX~\cite{Akerib:2013tjd, Akerib:2015rjg}. The direct detection constraint is obtained by comparing the spin-independent cross section for the scattering of $S$ off of a nucleon,
  \be
 \sigma_{\rm SI} = \frac{a_2^2 m_N^4 f^2}{\pi m_S^2 m_h^4} \, ,
 \ee
to the limits on $\sigma_{\rm SI}$ provided by LUX.
Here $m_N$ is the nucleon mass, and $f$ is the form factor that we take to be 1/3~\cite{Farina:2009ez, Giedt:2009mr}.
For values of $m_S$ larger than $m_h/2$ both the annihilation cross section $\sigma$, that enters the relic density calculation [see eq.~\eqref{eq:Z}], and the direct detection cross section $\sigma_{\rm SI}$ scale as $\frac{a_2^2}{m_S^2}$. However $\sigma$ is constant, with $\langle \sigma v_{\rm rel} \rangle \sim 3\times 10^{-26}$ cm$^{3}$ s$^{-1}$, while the direct detection constraint on $\sigma_{\rm SI}$ gets weaker at high masses as the number density of DM scales as $1/m_S$. That is why for $m_S > 110$ GeV the LUX constraint becomes too weak to exclude $S$ as a dark matter candidate. Other studies~\cite{deSimone:2014pda, Cline:2013gha} found a similar allowed window. The only hope to close the available high-mass window is with better sensitivity of future direct detection experiments.
  
At small values of $m_S$ the constraint from invisible Higgs decays becomes important. 
 We utilize here the results of a recent study~\cite{Zhou:2014dba} according to which the branching fraction of the Higgs 
 boson to invisible particles, in our case $h \to SS$, has to be less than 0.40 at 95\% confidence level.

  \begin{figure}[t]
\centering
\includegraphics[width=\textwidth]{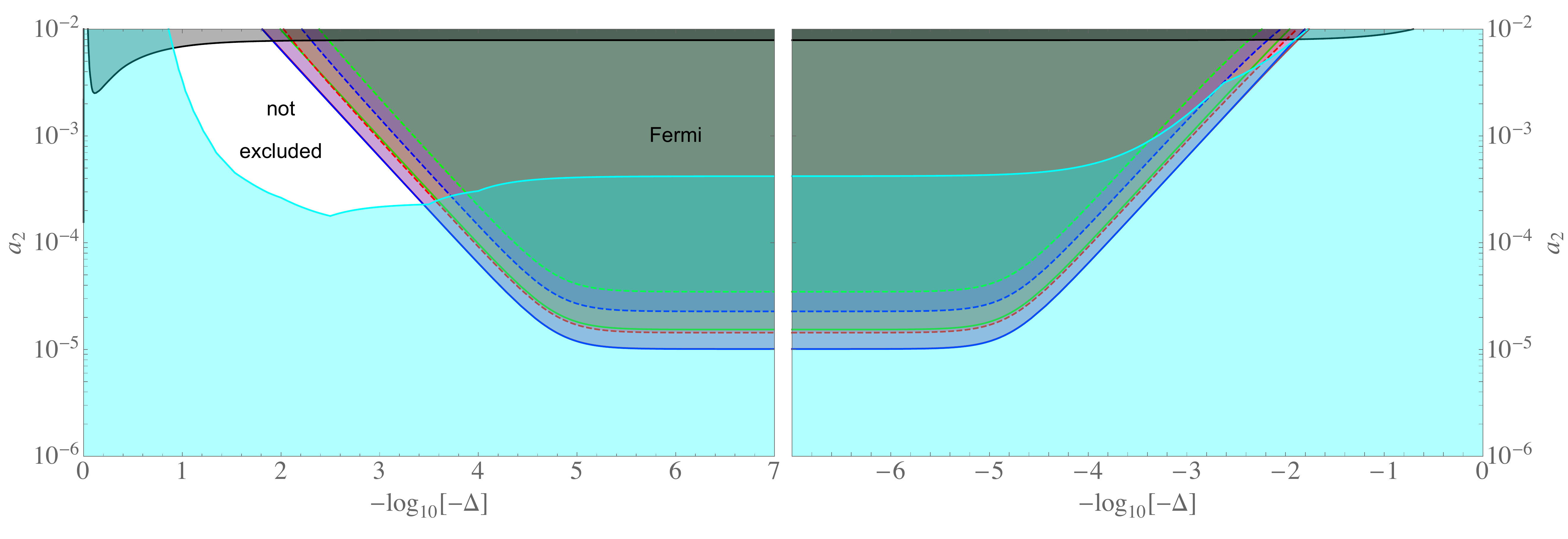}
\caption{We use the same lines and color code as in the previous figure, but we trade the mass $m_S$ for the dimensionless parameter $ \Delta \equiv \frac{2 m_S - m_h}{m_h}$ in the horizontal axis, in order to zoom into the resonant region. Here the Fermi constraints
are more severe than the ones from LUX (black line). The only region that escapes all current experimental constraints is
for $-10^{-0.9} < \Delta < -10^{-3.2}$, which corresponds to the mass range
54.9  GeV $< m_S <$ 62.8 GeV.
}
\label{fig:resonantregion}
\end{figure}

  The only limited region that escapes the above mentioned constraints is around $m_S \simeq m_h/2$, where 
  $m_h$ indicates the Higgs mass, a region we refer to as the ``{\em resonant}'' region. 
 In the resonant region the constraints from gamma-ray-line searches by Fermi~\cite{Ackermann:2013uma} are increasingly important.
  We indicate different Fermi constraints with different line coding in the figures. The 
plain lines correspond to the $SS \to \gamma \gamma$ channel, while the dashed lines to the $SS \to \gamma Z$ channel. We use different colors for different choices of the Galactic DM density profile: red is for Einasto,
blue for NFW, green for the Isothermal profile.
  
   Such constraints can be better appreciated by defining
  a new variable
  \be \label{eq:Delta}
  \Delta \equiv \frac{2 m_S - m_h}{m_h} \, .
  \ee
In Fig.~\ref{fig:resonantregion} we employ the variable $\Delta$ to show that the only region of the parameter space in which $S$ is still a viable
dark matter candidate is given by the range $-10^{-0.9} < \Delta < -10^{-3.2}$, which corresponds to the mass range
54.9  GeV $< m_S <$ 62.8 GeV. This range is slightly below the resonance, $m_S = m_h/2$, and the reason is simple: In the early universe, at temperatures close to the freeze-out, we cannot simply use the approximation that $S$ is non-relativistic, rather we should use eq.~\eqref{eq:GG} to compute the thermal averaged cross section, which is valid in all regimes. The 
kinetic energy of the particles $S$'s at that epoch, despite small, is non negligible and as a result the resonant condition in the annihilation, $s = m_h^2$, is met for values of $m_S$ smaller than
$m_h / 2$.  On the other hand, when we compute the Fermi constraints we are considering annihilations occurring
in the Galactic center today, in which case the temperature is much lower compared to the freeze-out temperature, the non-relativistic approximation is perfectly fine and the resonant condition is $m_S \simeq m_h/2$.

In our figure we assumed that for under-abundant thermal DM parameter space points (i.e. points where the relic density of DM is below the observed cosmological DM density) some other mechanism, such as non-thermal production or a modified cosmological history, has lead to the production of as much singlet scalar DM as the observed DM density. An alternate approach (recently pursued e.g. in Ref.~\cite{Cline:2013gha}) would have been to re-scale the singlet DM density according to the thermal relic density, i.e.
\be
\rho_{\rm singlet}=\rho_{\rm DM}\left(\Omega_{\rm singlet}/\Omega_{\rm DM}\right).
\ee
Such a rescaling would have significantly weakened both direct and indirect detection limits in the under-abundant regions. The rescaling is unnecessary on the cyan region indicating $\Omega_{\rm singlet}=\Omega_{\rm DM}$. This is the region that provides the maximal ranges for the two key parameters of the model, the mixing constant $a_2$ and the singlet mass $m_S$

Fig.~\ref{fig:resonantregion} elucidates the two key conclusions of our study: 
\begin{itemize}
\item[(1)] below the resonance, the singlet scalar DM model is only viable for singlet scalar masses in a narrow range, between 55 and 63 GeV, right below half the measured SU(2) Higgs mass, by direct detection from below and by indirect detection from above, and 
\item[(2)] the allowed range for the singlet-SU(2) mixing constant $a_2$ is constrained to between roughly $2\times10^{-4}\lesssim a_2\lesssim 7\times 10^{-3}$, by the relic density over-production constraint from below, and by direction detection from above.
\end{itemize}


\section{Discussion and Conclusions}

We have reassessed the real singlet scalar extension to the SM as a possible context for the explanation of the cosmological non-baryonic dark matter in light of the Higgs discovery and of improved direct and indirect dark matter detection constraints. We have demonstrated that two small regions of parameter space remain viable: (i) within the small mass range $55\lesssim m_S/{\rm GeV}\lesssim 63$ and for a similarly highly constrained range for the quartic coupling $a_2$ between the singlet and the SU(2) Higgs; (ii) for $m_S > 110$ GeV and a small range of $a_2$.  

A factor 20 improvement to the direct detection sensivity will conclusively test this model for $m_S$ below a TeV. Such an improvement is well within the reach of the planned G2 direct detection experiments SuperCDMS and LZ \cite{Cushman:2013zza}. New limits from Fermi-LAT will also shrink the  available parameter space, especially in the high-mass end of the currently open parameter space near the resonance. The high mass ($m_S > 110$ GeV) region is also still viable, albeit for a very small range of $a_2$, and can be further constrained essentially only with future direct detection experiments.

The singlet, real scalar dark matter model is a clear example of how minimal setups quickly become highly constrained, and thus highly predictive, with increasing quality of experimental data. Also, this specific context illustrates very clearly the complementarity across a variety of different dark matter detection strategies, including direct, indirect and collider searches. It will soon become clear whether or not this specific minimal extension to the Standard Model of particle physics is or not the culprit for the fundamental nature of dark matter.

\acknowledgments
We thank Alessandro Strumia for a constructive discussion about the high-mass region of this model. We are grateful to Nicolas Bernal and James Cline for useful discussions.
LF was supported by 973 Program of China under grant 2013CB837000 and National Natural Science of China under grant 11303096.
SP is partly supported by the US Department of Energy, Contract DE-SC0010107-001.

\appendix

\section{Annihilation cross sections} \label{sec:cross}
In this appendix we give explicit expressions for the cross sections relevant to the computation of the relic density and of the gamma-ray line constraints. The annihilation of $SS$ into any two-body final state $XX$, where $XX$ is either a pair of fermions or
a pair of gauge bosons, proceeds via the exchange
of a Higgs boson in the $s$-channel. The cross section times the relative velocity of the annihilating particles is
\be \label{eq:svgz}
(\sigma v_{\rm rel})_{XX}= \frac{8 a_2^2 v^2}{\sqrt{s}} | D_h (s) |^2 \Gamma_{h\to XX} (s) \, ,
\ee
where
\be
| D_h (s) |^2 \equiv \frac{1}{(s-m_h^2)^2 + m_h^2 \Gamma_h^2 (m_h)} \, .
\ee
The width in the above propagator is
\beqn
\Gamma_h (m_h) &=& \Gamma_{\rm vis} + \Gamma_{\rm inv} \, , \\
\Gamma_{\rm vis} &=& 4.07 \ {\rm MeV} \,  , \\
\Gamma_{\rm inv} &=& \frac{a_2^2 v^2}{8 \pi m_h} Re \sqrt{1-\frac{4m_S^2}{m_h^2}} \, ,
\eeqn
while each of the widths $\Gamma_{h\to XX} (s)$ in eq.~\eqref{eq:svgz} is obtained from the decay width of 
an off-shell Higgs into the $XX$ channel~\cite{Dittmaier:2011ti} substituting $(m_h^*)^2$ with $s$.
In the computation of the relic density we take into account all the possible SM two-body final states, including the $SS \to hh$ channel for which we adopt the same cross section as in Ref.~\cite{Cline:2013gha}:
\begin{align}
(\sigma v_{\rm rel})_{hh} =  \frac{a_2^2}{4 \pi s^2 V_S} & \left[  (a_R^2 + a_I^2) s V_S V_h +
8 a_2 v^2 \left(a_R - \frac{2 a_2 v^2}{s-2m_h^2} \right) \log \left\vert \frac{m_S^2 - t_+}{m_S^2 - t_-} \right\vert \right. \nn \\
& \left. + \frac{8 a_2^2 v^4 s V_S V_h}{(m_S^2 -t_-)(m_S^2 - t_+)} \right] \, ,
\end{align}	
where $V_i = \sqrt{1 - \frac{4m_i^2}{s}}$, $t_\pm = m_S^2 + m_h^2 - \frac{1}{2} s (1 \mp V_S V_h)$, and
\beqn
a_R &\equiv & 1+ 3m_h^2 (s-m_h^2) \vert D_h(s) \vert^2 \, , \\
a_I & \equiv & 3 m_h^2 \sqrt{s} \Gamma_h(m_h) \vert D_h(s) \vert^2 \, .
\eeqn

To apply the gamma-ray line constraints we are interested in the $\gamma \gamma$ and $\gamma Z$ channels. The 
corresponding widths are computed at one loop and are given by~\cite{Gunion:1989we, Djouadi:2005gi}
\beqn
 \Gamma_{h\to \gamma \gamma} (s) & = & \frac{\alpha^2 s^{3/2}}{256 \pi^3 v^2} \left\vert \sum_i N_{ci} e_i^2 F_i \right\vert ^2 \, , \\
 \Gamma_{h\to \gamma Z} (s) & = & \frac{\alpha m_W^2 s^{3/2}}{128 \pi^4 v^4}\left(1 - \frac{m_Z^2}{s}  \right)^3 \left\vert \sum_f \frac{N_{cf} e_f v_f}{c_W} A^H_f(\tau_f , \lambda_f) + A^H_W (\tau_W, \lambda_W)    \right\vert ^2 \, .
 \eeqn
 Here $\alpha$ is the electromagnetic fine structure constant, $c_W$ and $s_W$ are respectively the cosine and sine of the Weinberg angle, the index $i$ = $f, \ W$ identifies whether the particle running in the loop is a fermion or a $W$ boson, $N_{ci}$ is its color multiplicity, $e_i$ its electric charge in units of $e$, $\tau_i = 4m^2_i/s$ and $\lambda_i = 4m^2_i / m_Z^2$,
 \beqn \label{eq:Fs}
F_f&=& -2 \tau_f[1+(1-\tau_f)f(\tau_f)], \nn \\
F_W &=& 2+3\tau_W+3\tau_W(2-\tau_W)f(\tau_W),
\eeqn
\be \label{eq:ftau}
f(\tau)=\left\{ \begin{array}{ll}
\left[ \sin^{-1}(\sqrt{1/\tau})\right]^2, & {\rm if} \quad \tau\geq 1 \\
-\frac{1}{4}\left[\ln\left(\frac{1+\sqrt{1-\tau}}{1-\sqrt{1-\tau}}\right)-i \pi \right]^2, & {\rm if} \quad \tau < 1
\end{array} \right. \, \ ;
\ee
$v_f = 2 I_{3f} - 4 e_f s_W^2$, with $I_{3f}$ the fermion weak isospin,
\beqn
A^H_f &=& [I_1 (\tau, \lambda) - I_2 (\tau, \lambda)] \, \\
A^H_W (\tau, \lambda) &=& c_W \left\{ 4 \left(3 - \frac{s_W^2}{c_W^2} \right) I_2(\tau, \lambda) + \left[ \left(1 + \frac{2}{\tau} \right)\frac{s_W^2}{c_W^2} - \left(5+\frac{2}{\tau} \right) \right] I_1 (\tau, \lambda) \right\} \, ,
\eeqn
the functions $I_1$ and $I_2$ are given by
\beqn
I_1(\tau, \lambda) &=& \frac{\tau \lambda}{2(\tau - \lambda)} + \frac{\tau^2 \lambda^2}{2(\tau - \lambda)^2} \left[f(\tau) - f(\lambda) \right] + \frac{\tau^2 \lambda}{(\tau - \lambda)^2} \left[ g(\tau) - g(\lambda) \right] \, , \\
I_2(\tau, \lambda) & = & -\frac{\tau \lambda}{2(\tau - \lambda)} \left[f(\tau) - f(\lambda) \right] \, ,
\eeqn
with
\be \label{eq:gtau}
g(\tau)=\left\{ \begin{array}{ll}
\sqrt{\tau -1} \sin^{-1}(\sqrt{1/\tau}), & {\rm if} \quad \tau\geq 1 \\
\frac{\sqrt{1-\tau}}{2}\left[\ln\left(\frac{1+\sqrt{1-\tau}}{1-\sqrt{1-\tau}}\right)-i \pi \right], & {\rm if} \quad \tau < 1
\end{array} \right. \, \ .
\ee

\section{Note on the Fermi constraints} \label{sec:Fermi}
 In the model we are considering there are two annihilation processes that can give rise to gamma-ray lines. One is
 $SS \to \gamma \gamma$, the other $SS \to \gamma Z$. The corresponding cross sections are given in Appendix~\ref{sec:cross}. The Fermi collaboration provides limits~\cite{Ackermann:2013uma} on the flux, $\Phi_\gamma$,
of gamma rays from dark matter annihilation for photon energies between 5 and 300 GeV. They also translate
the limits on the flux directly into limits on the annihilation cross section to $\gamma \gamma$, thus comparing the cross 
section in our model for that channel to those constraints is straightforward. They leave to us the simple exercise 
of translating the flux limits into limits on the annihilation cross section to $\gamma Z$. The exercise is done as follows.
In the process $SS \to \gamma Z$, the energy of the monochromatic photon is 
\be \label{eq:Egamma}
E_\gamma = m_S \left( 1 - \frac{m^2_Z}{4 m_S^2} \right) \, .
\ee
As the range of the Fermi search starts at $E_\gamma = 5$ GeV, this implies that the minimum mass probed in this
channel is $m_S \simeq 48$ GeV, as it is reflected in the plot of Fig.~\ref{fig:all}. The flux is given by
\be \label{eq:flux}
\Phi_\gamma = \frac{\langle \sigma v_{\rm rel} \rangle_{\gamma Z}}{8 \pi m_S^2} J_{\rm ann} \, ,
\ee
where the J-factor, $J_{\rm ann}$, is the integral of $\rho(r)^2$ along the line of sight, with $\rho(r)$ the dark matter density profile. The J-factors corresponding to four different dark matter profiles are listed in Table III of Ref.~\cite{Ackermann:2013uma}. Combining eqs.~\eqref{eq:Egamma} and \eqref{eq:flux} we get
\be
\langle \sigma v_{\rm rel} \rangle_{\gamma Z} = \frac{1}{J_{\rm ann}} 8\pi \frac{1}{4} \left( E_\gamma + \sqrt{E_\gamma^2 + m_Z^2} \right)^2 \Phi_\gamma \, .
\ee
Then from the upper limits on $\Phi_\gamma$ listed in Table VII of Ref.~\cite{Ackermann:2013uma}, we can set constraints on $\langle \sigma v_{\rm rel} \rangle_{\gamma Z}$.

\bibliographystyle{JHEP}
\bibliography{SingletScalar}

\end{document}